# Monoclinic distortion in hyperhoneycomb Kitaev material β-ZnIrO$_3$ revealed by improved sample quality

Yuya Haraguchi, *[1] and Hiroko Aruga Katori[1]

[1]Graduate School of Engineering, Tokyo University of Agriculture and Technology, Koganei, Tokyo 184-8588

E-mail: *chiyuya3@go.tuat.ac.jp

The sample quality of the hyperhoneycomb lattice Kitaev magnet β-ZnIrO$_3$ was successfully improved by removing the maximum amount of moisture from the reaction ampoule. The X-ray diffraction structural analysis of the high-quality sample confirmed the presence of $P2_1/c$ superlattice peaks of the $Fddd$ structure in the original structural model. These peaks could not be distinguished due to the presence of impurities in the low-quality sample in a previous study. The structural analysis based on this monoclinic crystal structure model showed no chemical disorder, suggesting that the observed spin liquid type behavior is an intrinsic property unrelated to bond randomness. The details of the β-ZnIrO$_3$ structure revealed in this study will stimulate the further investigation of Kitaev physics.

The Kitaev model—a model recognized by Ising-like bond-dependent interactions on a tricoordinated lattice—is the most promising many-body model for the realization of quantum spin liquids (QSLs) characterized by fractionalized excitations [1,2]. Although more than 15 years have passed since the appearance of the Kitaev model, no material has realized a pure Kitaev model [3-5]. The QSL state in the Kitaev model suddenly collapses by additional nonKitaev interactions such as Heisenberg $J$ and off-diagonal interactions $\Gamma$ [6-8]. Almost all Kitaev candidates exhibit long-range magnetic order. The existence of the long-range magnetic order indicates that these Kitaev candidates have finite $J$ and $\Gamma$.

The deviation from the ideal model is manifested in reality as the degree of suppression of the magnetic order. Thus, physicists have discovered the possibility of reducing the absolute values of $J$ and $\Gamma$ and increasing the same of Kitaev interaction K. Winter et al. theoretically predicted that the magnitudes of $J$, $K$, and $\Gamma$ strongly depend on Ir–O–Ir bond angles $\varphi$ [9]. According to Winter's theory, strong ferromagnetic $K$ with negligible $J$ and $\Gamma$ is realized near $\varphi$ ~100°. This theoretical result strongly suggests materializing the Kitaev limit in realistic compounds by controlling the local crystal structure. Therefore, the worldwide search for new Kitaev candidates has been accelerating [3-5, 10, 11].

We have recently reported the successful synthesis of a new hyperhoneycomb lattice Kitaev magnet β-ZnIrO$_3$ with β-Li$_2$IrO$_3$ as a precursor using ion-exchange reactions at low temperatures [12]. In a previous study, we performed the structural analysis of the same space group

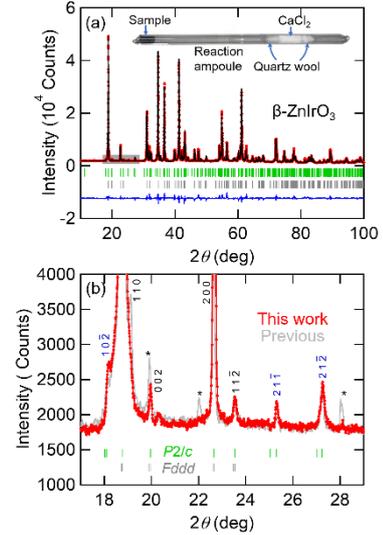

**Figure 1**. (a) The result of the Rietveld refinement method. The observed intensities (red circles), calculated intensities (black line), and their differences (blue curve at the bottom) are shown. The green and gray vertical bars indicate the positions of Bragg reflections of $P2_1/c$ and $Fddd$ structures, respectively. The inset shows the reaction ampoule. The area where calcium chloride is set protrudes from the furnace trapping a small amount of water from the starting material. (b) The enlarged view of the shaded area of Fig. 1(a). The numbers near the peaks indicate the index of the $P2_1/c$ structure. The blue numbers indicate the $P2_1/c$ superlattice peak in the $Fddd$ structure. For comparison, the XRD profile of the previous sample (gray data) [12] are shown behind the present data (red data) with unindexable peak positions marked by asterisk, evidencing an improvement in the sample quality.

$Fddd$ as the precursor. However, the XRD profile incorporated the Zn site disorder, and the analysis did not converge as expected. The specific heat and magnetic susceptibility at low temperatures exhibit minimal dependence on the magnetic field in the case of β-ZnIrO$_3$. This behavior is in contrast to the pronounced field dependence of specific heat and magnetic susceptibility observed in chemically disordered Kitaev candidates such as H$_3$LiIr$_2$O$_6$ [13] and Cu$_2$IrO$_3$ [14]. The theoretical investigation employing the bond-disordered Kitaev model effectively elucidates the observed magnetic field dependences [15]. Consequently, the available evidence

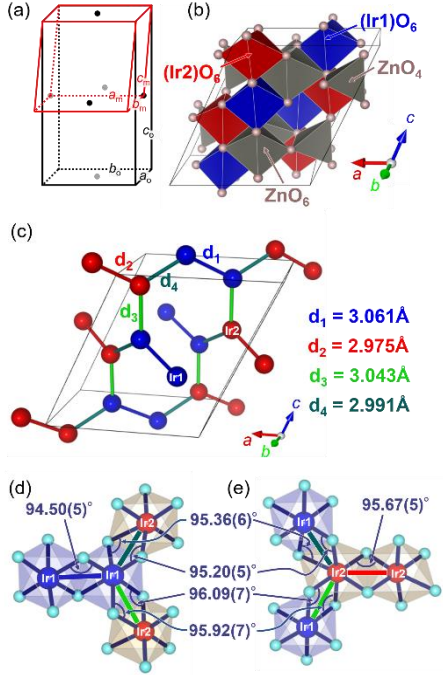

**Figure 2.** (a) The relation between orthorhombic (black) and monoclinic cells (red) is $a_m = b_o$, $b_m = -a_o$, and $c_m = -a_o/2 - c_o/2$. The closed circles indicate the position of the face center of each face in the orthorhombic cell. (b) The crystal structure of β-ZnIrO$_3$ with the $P2_1/c$ structure. (c) The hyperhoneycomb network, where two Ir-site and four Ir–Ir bonds with different lengths are shown differently. The local lattice network of IrO$_6$ octahedra of β-ZnIrO$_3$ around (d) Ir1 and (e) Ir2 sites with six different superexchange pathways with different Ir–O–Ir angles and four Ir–Ir distances.

**Table 1.** Renovated crystallographic parameters of β-ZnIrO$_3$ (space group: $P2_1/c$) determined from powder X-ray diffraction experiments. The obtained lattice parameters are $a = 8.7690(2)$ Å, $b = 5.9313(2)$ Å, $c = 9.9355(3)$ Å, and $\beta = 116.1382(2)°$. $B$ is the atomic displacement parameter. $g$ is the occupancy factor.

| atom | site | g | x | y | z | B (Å$^2$) |
|---|---|---|---|---|---|---|
| Ir1 | 4e | 1 | 0.329 44(7) | 0.123 33(32) | 0.426 39(8) | 0.92(2) |
| Ir2 | 4e | 1 | 0.1594(4) | 0.1264(13) | 0.0793(2) | 0.42(2) |
| O1 | 4e | 1 | 0.0726(7) | 0.606(3) | 0.411(1) | 0.4(2) |
| O2 | 4e | 1 | 0.0983(8) | 0.128(4) | 0.4341(16) | 0.57(5) |
| O3 | 4e | 1 | 0.3947(7) | 0.148(3) | 0.0950(9) | 0.7(3) |
| O4 | 4e | 1 | 0.5628(7) | 0.113(2) | 0.4092(10) | 1.0(2) |
| O5 | 4e | 1 | 0.7422(7) | 0.3949(12) | 0.2525(11) | 0.40(15) |
| O6 | 4e | 1 | 0.2557(9) | 0.3543(10) | 0.2515(15) | 0.15(6) |
| Zn1 | 4e | 1 | 0.500 93(18) | 0.3963(6) | 0.265 88(16) | 0.53(3) |
| Zn2 | 4e | 1 | 0.122 55(14) | 0.6259(5) | 0.2556(2) | 0.35(2) |

A polycrystalline β-ZnIrO$_3$ sample was prepared based on a previously reported method. A molar ratio of pelletized β-Li$_2$IrO$_3$: ZnSO$_4$: KCl = 1: 15: 8 was allowed to react under a vacuum atmosphere at 400 °C for 100 h [12]. The following methods effectively improved the sample quality. Calcium chloride was set in the cooling section of the reaction ampoule opposite to the sample side to trap the trace amount of moisture present in the starting materials, as shown in the inset of Fig. 1(a). This technique is based on the calcium chloride tube method of organic synthetic chemistry. This improvement suggested that a trace amount of moisture, possibly present in the starting materials, decomposed the reactants. The product was characterized by powder X-ray diffraction (XRD) experiments using a diffractometer with Cu–Kα radiation. The cell parameters and crystal structure were refined by the Rietveld method using the Z-Rietveld software [16].

Figure 1(a) shows the observed XRD pattern with Bragg positions based on $P2_1/c$ (green) and $Fddd$ (gray) structures for comparison. Peaks that the $Fddd$ structure cannot index are superlattice peaks indexable by the $P2_1/c$ structure with the lattice constants $a = 8.7690(2)$ Å, $b = 5.9313(2)$ Å, $c = 9.9355(3)$ Å, and $\beta = 116.1382(2)°$. This structure is similar to the 1.5 GPa high-pressure phase of β-Li$_2$IrO$_3$ [17]. An enlarged view of the shaded area in Figure 1(a) is shown in Figure 1(b). The refined crystallographic parameters are listed in Table 1. The superlattice reflections of the $Fddd$ structure can be indexed as the primitive monoclinic unit cell with a $P2_1/c$ space group. As a comparison, the XRD spectrum of the previous sample [12] shows impurity peaks that cannot be indexed by $P2_1/c$. Therefore, the improved sample quality allows us to distinguish the peak representing impurities from the $P2_1/c$ superlattice. The relation between the orthorhombic and monoclinic cells is $a_m = b_o$, $b_m = -a_o$, and $c_m = -a_o/2 - c_o/2$ [Figure 2(a)]. An initial model of the $P2_1/c$ structure is constructed by reducing some symmetry from the $Fddd$ structure. As shown in Figure 1(a), the Rietveld refinement converges better with the $P2_1/c$ structure depicted in Figures 2(b–c). Zn ions occupy 6- and 4- coordinated sites similar to the $Fddd$ structure. The structural strongly implies that the spin liquid-like behavior observed in β-ZnIrO$_3$ is not predominantly attributable to the chemical disorder. These observations are inconsistent with the chemical disorder in $Fddd$ structure-based structural analysis. We proposed a hidden mechanism that protected QSL states from chemical disorders in the previous study. However, the mechanism is imperative in the presence of a chemical disorder and nonimperative in its absence. This problem requires a more comprehensive structural model through a more detailed structural analysis to reveal the presence or absence of chemical disorder.

In this letter, we report that all the small extra XRD peaks, considered impurities, can be indexed by a $P2_1/c$ superlattice of $Fddd$ by further improving the sample quality. The purity of the sample was also confirmed through magnetization measurements, revealing a drastic reduction in glass contribution that was present in the previous sample, indicating that glass anomaly was not intrinsic. Based on this finding, we refined the structural analysis and improved the crystal structure without any disorder of Zn sites.

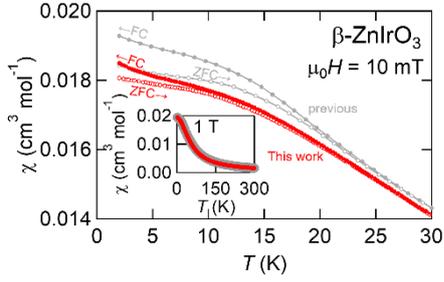

**Figure 3**. Comparison of magnetic susceptibility of the previous and new samples. The measurements were conducted upon heating after zero field cooling and then cooling under the magnetic field of 10 mT. The inset shows the temperature dependence of the magnetic susceptibility from 2 to 300 K under the magnetic field of 1 T.

analysis with the *Fddd* structure model requires a decrease in the occupancy of Zn sites, whereas the structural analysis with the $P2_1/c$ structure model shows that the occupancy of Zn sites both converges entirely to 1. This analysis evidences that the observed QSL-like behavior is intrinsic and not derived from chemical disorder. The hypothetical β-value in the *Fddd* structure with the angle between the $-a_o$ ($= b_m$) and $-a_o/2 - c_o/2$ ($= c_m$) vectors is estimated as β = 118.39°. In contrast, the refined β value is 116.14°, a slight distortion of Δβ ~ 2° from the *Fddd* structure.

The two-site $Ir^{4+}$ ions form a distorted hyperhoneycomb lattice with four types of Ir–Ir bonds. There are six superexchange interactions via oxygen ions [Figures 2(d–e)], the magnitude of which should depend on the Ir–O–Ir bond angle. The bond angles are in the range of 94.5–96.1°. This slight variation in bond angles indicates that the three bond interactions at each site on the spin model are almost equivalent.

The temperature dependence of magnetic susceptibility was measured under a low magnetic field of 10 mT to compare the quality of the previous and new samples, as illustrated in Figure 3. The χ-data of the previous sample reveals a slight glassy behavior at ~18 K. It is challenging to discern whether this anomaly is intrinsic or caused by extrinsic factors. In contrast, this glass-like magnetic anomaly is also significantly suppressed in the new sample with significant removal of impurities. The comprehensive temperature-dependent magnetic susceptibility is replicated, as shown in the inset of Figure 3. Thus, the fundamental magnetic properties exhibit minimal deviation. These observations suggest that glass contribution is probably attributable to extrinsic factors, such as the trace amounts of ferromagnetic impurities. This provides strong evidence of significantly improved sample quality of the new samples.

According to the theoretical calculations using nonperturbative exact diagonalization by Winter *et al.*, the magnitude of these interactions ($K$, $J$, $\Gamma$) strongly depends on the Ir–O–Ir bond angle [9]. The Heisenberg interaction is weak based on this relationship. Moreover, antiferromagnetic $\Gamma$ is comparable to ferromagnetic $K$. Since it has been theoretically shown that $\Gamma$ appears as an anisotropy of the Weiss temperature, it is necessary to establish a single-crystal growth method to clarify the details of $\Gamma$ contribution in β-$ZnIrO_3$.

An improved sample quality enabled us to refine the structure of β-$ZnIrO_3$ and confirm that it forms a monoclinically-distorted hyperhoneycomb lattice without the chemical disorder. However, the spin model predicted by this structure does not fully account for the observed spin liquid-type behavior. The comparable $\Gamma$ and Kitaev interactions should disrupt the Kitaev spin liquid and stabilize the magnetic order. Distortion to a monoclinic structure also complicates the spin model by introducing two Ir sites, four Ir–Ir bonds, and six Ir–O–Ir superexchange pathways. According to calculations by Stavropoulos et al., $\Gamma$ interaction also induces a counter-rotating spiral order in the 3D Kitaev model [18]. Given that spiral magnetism is typically the result of the competition between several magnetic interactions, the magnetic ordering temperature is significantly suppressed with the interaction energy [19-22]. Therefore, magnetization measurements performed up to 2 K in the present stage may not accurately capture the magnetic ordering in this system. Magnetization measurements at cryogenic temperatures using dilution coolers are planned to clarify the reasonable spin model in β-$ZnIrO_3$. Nevertheless, there is still much room for the theoretical development of the magnetic phase diagram of the ground state in the three-dimensional Kitaev-$\Gamma$ model, which will be clarified theoretically in the future.

In summary, we have successfully improved the quality of our hyperhoneycomb iridium oxide β-$ZnIrO_3$ samples. Furthermore, the structural analyses of high-quality β-$ZnIrO_3$ samples revealed the materialization of a chemical disorder-free monoclinic $P2_1/c$ structure. These results demonstrate that the spin liquid type behavior of β-$ZnIrO_3$ is intrinsic and not mimicked by bond randomness.

This work was supported by the Japan Society for the Promotion of Science (JSPS) KAKENHI Grant Number JP22K14002 and JP21K03441. Part of this work was performed through joint research at the Institute for Solid State Physics, the University of Tokyo.